\documentclass[11pt,twoside]{book}
\usepackage{konkolyproc2}
\usepackage{longtable}
\usepackage{amsmath,amssymb}
\usepackage{graphicx}
\usepackage{lscape}
\usepackage{index}
\usepackage{natbib}
\usepackage{bigdelim}
\usepackage{multirow}
\makeindex

\begin{document}

\pagestyle{myheadings}
\setcounter{equation}{0}\setcounter{figure}{0}\setcounter{footnote}{0}\setcounter{section}{0}\setcounter{table}{0}\setcounter{page}{1}
\markboth{Szabados, Szab\'o \& Kinemuchi}{RRL2015 Conf. Papers}
\title{The first decade of RR Lyrae space photometric observations}
\author{L\'aszl\'o Moln\'ar }
\affil{Konkoly Observatory, MTA CSFK, Budapest, Hungary}

\begin{abstract}
Space-based photometric telescopes stirred up stellar astrophysics in the last decade, and RR Lyrae stars have not been an exception from that either. The long, quasi-continuous, high-precision data from MOST, CoRoT and \textit{Kepler} revealed a wealth of new insights about this well-known variable class. One of the most surprising mysteries turned out to be the apparent omnipresence of a common additional mode in all RRd and RRc stars. Moreover, fundamental-mode stars seem to populate two distinct classes, one of which is characterized by the presence of additional modes and/or modulation, and another limited to strict single-mode pulsation. The presence of additional modes and multiple modulations in RRab stars allowed us to construct Petersen diagrams for these parameters: while the pulsation modes show clear structures according to period ratios, there seems to be no relation between the modulation periods themselves.
\end{abstract}

\section{Double-mode stars and the rise of the $f_X$ mode}
The era of space-based photometry of RR Lyrae stars started a decade ago, when the MOST space telescope observed the double-mode star AQ~Leo \citep{Gruberbauer2007}. Since the space photometric revolution started with an RRd star, it is fitting to start this review with the classical double-mode stars. Given that the double-mode region of the instability strip is very narrow (see, e.g.\ \citealt{Szabo2004}), RRd stars are inherently rare, and only a few have been observed from space so far. MOST and CoRoT observed one-one each: \citet{Gruberbauer2007} and \citet{Chadid2012} both identified various additional modes in AQ~Leo and CoRoT ID 0101368812, respectively. 

The original \textit{Kepler} field did not contain any known RRd stars, but the first K2 campaigns already presented us with three targets \citep{kurtz2015,molnar2015a}. Despite the diverse notations used in these papers for the additional modes, the comparison of the frequency content of these stars reveal a strikingly similar pattern. Instead of a collection of various additional modes, all five stars exhibit the same $f_X$ or $f_{\it 0.61}$ mode that was originally identified in first-overtone Cepheids and RR Lyrae stars (\citealt{moskalik2014a}, and references therein). Although the space-based RRd sample is still small, the data so far indicate that the $f_X$ mode and thus the triple-mode state is common (and quite possibly universal) within this subclass.

\section{First-overtone stars: the new double-mode class}
\begin{figure}[!ht]
\includegraphics[width=1.0\textwidth]{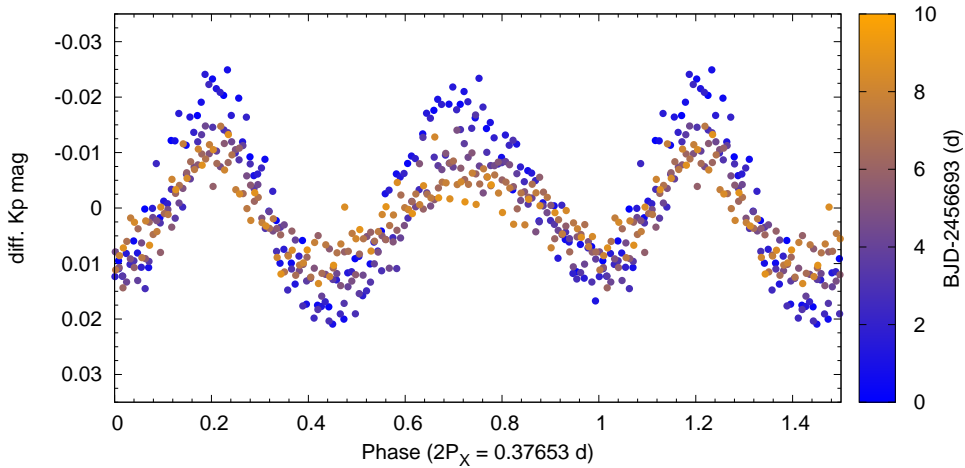}
\caption{The phase diagram of the disentangled $f_X$ mode in the RRc star EPIC 60018224 \citep{molnar2015a}.The frequency components of the first overtone and the combination terms were removed. The data is folded with $P = 2/f_X= 1/f_\ell$. The color coding shows the temporal evolution of the mode. The amplitude decreased significantly even during the short K2-E2 run. Also note that the two minima are not equidistant from each other.    } 
\label{molnar-fig1} 
\end{figure}

The mysterious $f_X$ mode was identified in almost all first-overtone stars observed from space, with the notable exception of the modulated star CSS J235742.1--015022 \citep{szabo2014,molnar2015a,moskalik2015}. Although the space-based sample is also small for RRc stars, with only 4 \textit{Kepler},  4 K2, and 2 CoRoT targets published so far, other observations also point towards the ubiquity of this additional mode. Notably, the ground-based data from the OGLE surveys and an intensive monitoring of M3 provided us with a large variety of high-precision RRc light curves \citep{jurcsik2015,netzel2015a,netzel2015c}. These studies revealed that the $f_X$ mode splits into three sequences between period ratios $P_X/P_1 \sim$ 0.60--0.64, similarly to the three sequences observed in Cepheids. Half-integer frequencies at n/2 $f_X$ values often appear, and all these frequency components show strong amplitude and phase variations on various timescales. Half-integer peaks may indicate that the mode is period-doubled, e.g. two cycles with different amplitudes (or shapes) alternate in time. Both effects can be observed in the phase diagram of an $f_X$ mode shown in Figure \ref{molnar-fig1}. 

The origin of this (or these) modes is still an open question. If we accept $f_X$ as the intrinsic pulsation frequency, the corresponding non-radial modes in this regime experience either very strong damping or very strong cancellation due to high spherical degrees required (up to $\ell \sim 40-50$), or both. An alternative hypothesis is that the pulsation frequency is in fact $f_X/2$, originating from $\ell = 8$ and 9 modes and then all the other peaks are simply the $nf_\ell$ harmonics (Dziembowski \& Smolec, in prep., also in these proceedings). In that case, period doubling does not occur, but a mode geometry is required where the observed amplitude of the $f_X = 2f_\ell$ can be much higher than that of $f_\ell$.  

Space-based photometry also revealed that not only the $f_X$ mode but the first overtone itself may exhibit temporal variations. The four \textit{Kepler} targets showed fluctuations in amplitude and phase on very different timescales, but interestingly these timescales are the same for $f_1$ and $f_X$ for any given star, indicating some connection, such as mode interaction between them \citep{moskalik2015}. We note that these variations have small amplitudes and show irregularities, so they appear to be different from the Blazhko effect observed in RRc stars.

Finally, not all RRc stars are limited to the first overtone and one or more $f_X$-type mode. Yet another mode was discovered in one of the \textit{Kepler} stars and later in some OGLE targets \citep{moskalik2014a,netzel2015b}. This mode has a period ratio of $P_1/P_{0.68} = 0.686$, e.g. the mode period is longer than the first overtone, and even longer than the scaled period of the fundamental mode with $P_0/P_{0.68}\sim 0.92$. At the moment the origin of this mode is even less clear than that of the $f_{0.61}$ mode. 

\section{RRab stars: a group of two classes}
Most of the RR Lyrae stars pulsate in the fundamental mode, so naturally most of the targets of space photometric missions were RRab stars as well. These included the eponym, RR Lyr that was being observed for almost four years continuously \citep{kolenberg2011}. One expectation about RR Lyr was that the \textit{Kepler} data would reveal the details of the suspected four-year cycle and the associated modulation phase shift in the star \citep{detreszeidl1973}. What transpired, instead, was that the levels of both amplitude and phase modulation were decreasing steadily over the four years. Although the amplitude modulation seemed to recover towards the end of the data, the phase variation continued to decrease even afterwards, according to ground-based measurements \citep{leborgne2014}. 

\begin{figure}[!ht]
\includegraphics[width=1.0\textwidth]{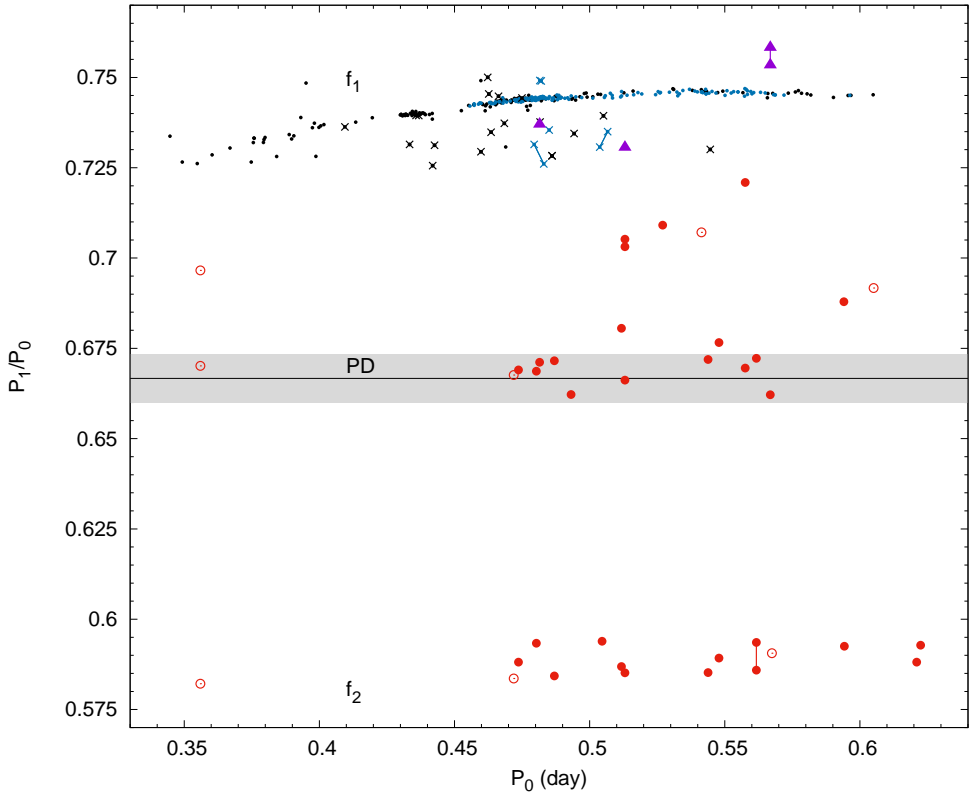}
\caption{Petersen diagram of the low-amplitude additional modes observed in the CoRoT (empty symbols) and \textit{Kepler}/K2 (filled symbols) RRab stars. Triangles correspond the three instances where the first overtone was potentially detected. Black and blue dots mark the position of classical RRd stars from the OGLE and CSS catalogs. Black crosses indicate the modulated or mode-switching bulge RRd stars, while blue crosses are the modulated RRd stars in M3. The black horizontal line marks the $P_1/P_0=2/3$ ratio, the grey area is the $< \pm 1\%$ difference. } 
\label{molnar-fig2} 
\end{figure}

Although \textit{Kepler} did not find the 4-year cycle, it revealed other, very important details about RR Lyr, and the entire variable class, with the discovery of period doubling and the presence of additional modes \citep{kolenberg2010,szabo2010}. Since then, these additional modes stars were found in several other RRab stars, although the amplitudes rarely exceed a few mmag. A Petersen-diagram (e.g.\ mode period ratios versus the period of the fundamental mode) of these new frequency components is presented in Figure \ref{molnar-fig2}. This plot shows all peaks found between $f_0$ and $2f_0$, including the ones close to $3/2 f_0$ half-integer peak (HIF). This component itself does not indicate a separate mode, just the presence of period doubling, but it often splits into a forest of peaks, with the dominant one shifted from the exact $3/2 f_0$ value (see, e.g. \citealt{benko2010}). The shift can be attributed to the facts that both the fundamental mode and period doubling itself is variable in time: this effect was also observed in BL Her hydrodynamic models \citep{smolecmoskalik2012}. However, the diagram indicates that the distinction between HIFs and nearby, bona-fide additional modes (or related peaks) towards the first overtone can be indeed subtle. In contrast, the $f_2$ mode below clearly separates from the rest of the peaks, and only appears at period ratios between $P/P_0 = 0.580-0.595$. 

The diagram also shows that the first overtone rarely occurs in RRab stars as an additional mode. The three examples so far are RR Lyr itself, V445 Lyr (KIC 6186029), and EPIC 206280713 (\citealt{guggenberger2012,molnar2012}, Juh\'asz et al., in these proceedings). If we compare the period ratios of these stars to the classical double-mode stars, we can see that RR Lyr is very peculiar, with a ratio of $P_1/P_0 > 0.75$. Interestingly, although the other two stars lie outside the main double-mode ridge as well, they perfectly overlap with the non-standard RRd stars such as modulated and mode-switching objects \citep{soszynski2014,jurcsik2015,smolec2015a,smolec2015b}. This further suggests that we indeed observe the first radial overtone is these stars. One of the great advantages of the step-and-stare approach of the K2 mission is that this Petersen diagram is going to be populated with at least one order of magnitude more objects that may reveal more structures and could help us to identify the various additional modes modulated RRab stars exhibit. Another important addition to this sample will be the OGLE data that certainly holds several examples of RRab stars with additional modes.  

Apart from these additional modes, there is at least one indication that RRab stars may also exhibit modes with (apparent) periods longer than the fundamental mode. A low-amplitude signal was tentatively identified at $P/P_0 = 0.695$ in EPIC 60018644 \citep{molnar2015a}. Although that detection was based only the short K2 Engineering data set, the star will fall on silicon during Campaign 12 of the K2 mission, making a much longer follow-up study possible.  

Space-based photometric observations also revealed a striking dichotomy among the fundamental-mode stars: those that are not modulated, do not contain additional modes either. Therefore non-Blazhko RRab stars seem to be the one and only subclass among the RR Lyraes that are truly monoperiodic \citep{szabo2015}. This dichotomy was further supported by \citet{benko2015}, who discovered low-level modulations in two $Kepler$ stars that initially appeared to have additional modes without the Blazkho effect. The existence of years-long, sub-mmag level modulation, as in the case of KIC 7021124, represents an important challenge in the analysis of the shorter K2 and TESS observations. The findings from CoRoT and $Kepler$ suggest that in ambiguous cases the presence of additional modes or period doubling can be considered as proxy for the presence of a long-term and/or low-amplitude Blazhko effect. 

\section{The Blazhko effect: new insights, new models}
\begin{figure}[!ht]
\includegraphics[width=1.0\textwidth]{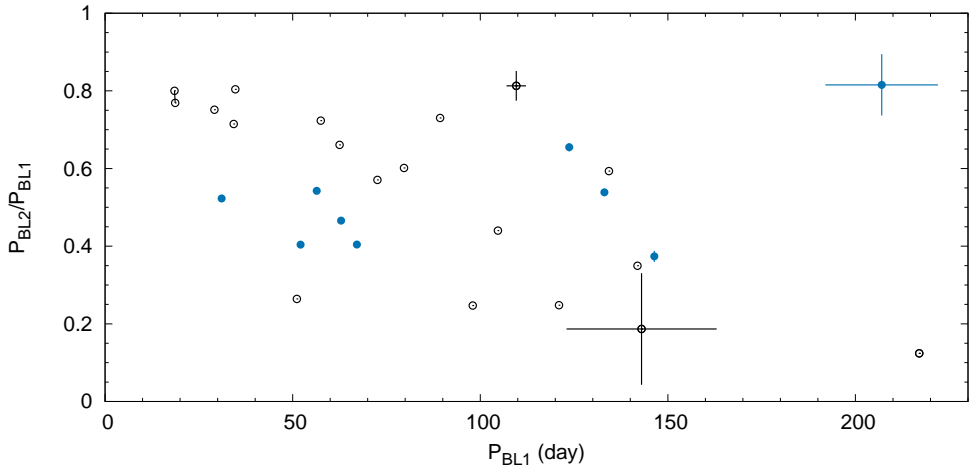}
\caption{Petersen diagram of the Blazhko-RRab stars where two modulation periods can be determined. Filled points represent the \textit{Kepler} data, empty circles are ground-based measurements. } 
\label{molnar-fig3} 
\end{figure}

Beside these important details of the pulsation, the long-term, continuous coverage of the space missions also showed us the bigger picture: the true shape of the Blazhko effect. The long observations of the original \textit{Kepler} mission were especially useful to investigate the temporal behavior of the modulation. \citet{benko2014} found that most of the Blazhko RRab stars in that sample show either multiperiodic or irregular modulation. Together with the rest of the known multiperiodic Blazhko stars, a Petersen diagram of modulation periods can be constructed. In Figure \ref{molnar-fig3}, I plotted all stars that have both modulation periods determined, and assigned the two as $P_{BL1} > P_{BL2}$, regardless of the amplitude ratios. Properties of Blazhko stars are collected in the database of \citet{skarka2013}\footnote{\texttt{http://www.physics.muni.cz/$\sim$blasgalf/}}. The connected circles represent CZ Lac that changed both of its modulation periods over time \citep{sodor2011}. The figure, at least with the current sample, does not suggest any particular structure or relation between the modulation periods.

\section{Classification considerations}
The new, intriguing details that space-based photometry have been providing us with may prompt us to redefine, or at least reconsider the old Bailey-type classification scheme. The picture that have emerged in the last few years can be summarized as follows:

\begin{itemize}
\item \textbf{RRd class:} stars where both the fundamental mode and the first overtone pulsate with large amplitudes (few tenths of a magnitude) and a low-amplitude $f_X$-type mode is also present. Blazhko subtype exists, both the fundamental mode and the first overtone can be modulated.
\item \textbf{RRc class:} stars where the first overtone pulsates with large amplitude (few tenths of a magnitude) and a low-amplitude $f_X$-type mode is also present. Blazhko subtype exists. 
\item \textbf{Blazhko RRab stars:} stars where the fundamental mode pulsates with large amplitudes (few tenths of a magnitude) and displays quasi- or multiperiodic modulation. Other low-amplitude modes and/or period doubling are usually present. 
\item \textbf{Non-Blazhko RRab stars:} stars where only the fundamental mode pulsates and it has a constant amplitude.
\end{itemize}

\section{Future prospects}
In the coming years, space-based photometry is going to provide us with an increasing amount of data about RR Lyrae stars. The K2 mission is expected to observe a few thousand targets, including a few hundred stars in the Sagittarius stream and the Galactic bulge in various campaigns (Plachy et al., in these proceedings).  Later on, the TESS space telescope will also cover at least hundreds, but possibly thousands of the nearby, brighter population of RR Lyraes. These observations will allow us to sample all various subtypes, and promises to fill up the Petersen diagram of RRab stars, for example. Data from various populations, such as the Sagittarius stream, can also reveal incidence rate differences in the Blazhko effect, at least for short-period modulations (Kov\'acs, in these proceedings).

Beside the new observations, the existing data sets still contain valuable information that can be extracted with various methods. The CoRoT data set is still not fully analyzed and it may contain additional RR Lyrae stars \citep{klagyivik2015}. The \texttt{Kepler} field, in comparison, is much more thoroughly explored but ingenious methods may allow us to extract even more. The smear measurements, an ancillary engineering data produced by the CCD modules that allows us to recover the light variations of the brightest stars in the field of view \citep{pope2015}. One result of this method is that the light curve of RR Lyr itself can be recovered for quarters Q3 and Q4 when the star was not part of the observing program \citep{kolenberg2011}. With the smear data, the photometry of RR Lyr now spans the whole mission from Q1 to Q17. Although smear is not very important for the generally fainter RR Lyraes, it will allow us to observe bright Cepheids, such as Y~Sgr or V350~Sgr in the future with K2, without assigning a large number of pixels to them. 

Another important source is the recently released RR Lyrae catalog of the PanSTARSS 3$\pi$ survey (\citealt{hernitschek2015}, also in these proceedings). Asteroid surveys like the Catalina and LINEAR sky surveys have avoided the vicinity of the Galactic plane, so we have had no reliable sources for fainter RR Lyraes in those areas. With the 3$\pi$ catalog we will be able to search for variables in the existing \textit{Kepler} and K2 observations as well. The original \textit{Kepler} field included two superstamps covering the two open clusters NGC 6791 and NGC 6819, where faint, previously unknown RR Lyraes may still hide. The large superstamps of the K2 mission, as well as the generous target pixel masks of the early campaigns are also prime targets for such a follow-up search, and could verify the purity measures of the catalog in various galactic directions. The $3\pi$ RR Lyrae catalog extends to about 21.5~mag in the PanSTARSS \textit{r} band. As the photometry of the variables in the galaxy Leo IV showed, \textit{Kepler} is perfectly capable to recover stars this faint during the span of one campaign \citep{molnar2015b}.  

The increasing influx of high-precision data presents new challenges for theoretical studies as well. The golden sample of space-based photometry, together with the large ground-based catalogs, can give important constraints on the structure and evolution of the Milky Way. Pulsation theory also faces new challenges with the appearance of the additional modes. While progress has been made for some of those, especially the origin of period doubling and related modes in RRab stars, and the new developments about the $f_X$ mode are also promising, several open questions remain. These include the rest of the additional modes, such as the origin of the $f_2$ mode in RRab stars, the nature of the long-period modes, the explanation for the mode-switching stars, and so on. And let us hope that the answers to these questions may also lead us closer to the solution of the Blazhko effect itself.

\section*{Acknowledgements}
 L.M.\ was supported by the J\'anos Bolyai Research Scholarship of the Hungarian Academy of Sciences. 
This research has been supported by the Lend\"ulet-2009 and LP2014-17 Program of the Hungarian Academy of Sciences,
and by the NKFIH PD-116175 grant of the Hungarian National Research, Development and Innovation Office. The research leading to these results has received funding from the European Community's Seventh Framework Programme (FP7/2007-2013) under grant agreements no. 269194 (IRSES/ASK) and no. 312844 (SPACEINN).


\begin{thebibliography}{}      
\bibitem[Benk\H{o} et~al.(2010)]{benko2010}Benk\H{o}, J. M., et al., 2010, MNRAS, 409, 1585
\bibitem[Benk\H{o} et~al.(2014)]{benko2014}Benk\H{o}, J. M., et al., 2014, ApJS, 213, 31
\bibitem[Benk\H{o} \& Szab\'o(2015)]{benko2015}Benk\H{o}, J. M., Szab\'o, R., 2015, ApJ, 809, L19
\bibitem[Chadid(2012)]{Chadid2012} Chadid, M., 2012, A\&A, 540, 68
\bibitem[Detre \& Szeidl(1973)]{detreszeidl1973}Detre, L., Szeidl, B. 1973, IBVS, 764, 1
 \bibitem[Gruberbauer et~al.(2007)]{Gruberbauer2007} Gruberbauer, M., et al., 2007, MNRAS, 379, 1498
 \bibitem[Guggenberger et~al.(2012)]{guggenberger2012} Guggenberger, E., et al., 2012, MNRAS, 424, 649
 \bibitem[Hernitschek et~al.(2015)]{hernitschek2015} Hernitschek, et~al., 2015, ApJS, accepted, arXiv:1511.05527
 \bibitem[Jurcsik et~al.(2015)]{jurcsik2015} Jurcsik, J., et~al., 2015, ApJS, 219, 25
 \bibitem[Klagyivik et~al.(2015)]{klagyivik2015} Klagyivik, P., et~al., 2015, AJ, accepted, arXiv:1510.01936
  \bibitem[Kolenberg et~al.(2010)]{kolenberg2010} Kolenberg, K., et al., 2010, ApJ, 713, L198, 
  \bibitem[Kolenberg et~al.(2011)]{kolenberg2011} Kolenberg, K., et al., 2011, MNRAS 411, 878 
 \bibitem[Kurtz et~al.(2015)]{kurtz2015} Kurtz, D.~W., Bowman, D.~M., Ebo, S.~J.,Moskalik, P., Handberg, R., Lund, M.~N., 2015, MNRAS, accepted, arXiv:1510.03347
 \bibitem[LaCluyz\'e et~al.(2004)]{lacluyze2014} LaCluyz\'e, A., et al., 2004, AJ, 127, 1653  
 \bibitem[Le Borgne et~al.(2014)]{leborgne2014} Le Borgne, J.~F., et al., 2014, MNRAS, 441, 1435  
 \bibitem[Moln\'ar et~al.(2012)]{molnar2012} Moln\'ar, L., et al., 2012, ApJ, 757, L13
 \bibitem[Moln\'ar et~al.(2015a)]{molnar2015a} Moln\'ar, L., et al., 2015a, MNRAS, 452, 4283 
 \bibitem[Moln\'ar et~al.(2015b)]{molnar2015b} Moln\'ar, L., et al., 2015b, ApJ, 812, 2 
\bibitem[Moskalik(2014)]{moskalik2014a} Moskalik, P., 2014, IAUS, 301, 249
\bibitem[Moskalik et~al.(2015)]{moskalik2015} Moskalik, P., et~al., 2015, MNRAS, 447, 2348
\bibitem[Netzel et~al.(2015a)]{netzel2015a} Netzel, H., et~al., 2015a, MNRAS, 447, 1173
\bibitem[Netzel et~al.(2015b)]{netzel2015b} Netzel, H., et~al., 2015b, MNRAS, 451, L25
\bibitem[Netzel et~al.(2015c)]{netzel2015c} Netzel, H., et~al., 2015c, MNRAS, 453, 2022
\bibitem[Pope et~al.(2015)]{pope2015} Pope, B. J S., et~al., 2015, MNRAS, 455, 36
\bibitem[Skarka(2013)]{skarka2013} Skarka, M., 2013, A\&A, 549, 101
\bibitem[Smolec \& Moskalik (2012)]{smolecmoskalik2012} Smolec, R., Moskalik, P., 2012, MNRAS, 426, 108
\bibitem[Smolec et~al.(2015a)]{smolec2015a} Smolec, R., et al., 2015a, MNRAS, 447, 3756
\bibitem[Smolec et~al.(2015b)]{smolec2015b} Smolec, R., et al., 2015b, MNRAS, 447, 3847
\bibitem[S\'odor et~al.(2011)]{sodor2011} S\'odor, \'A., et~al., 2011, MNRAS, 411, 1585
\bibitem[Soszy\'nski et~al.(2014)]{soszynski2014} Soszy\'nski, I., et al., 2015, AcA, 64, 177
\bibitem[Szab\'o et~al.(2004)]{Szabo2004} Szab\'o, R., Koll\'ath, Z., Buchler, J.~R., 2004, A\&A, 425, 627
\bibitem[Szab\'o et~al.(2010)]{szabo2010} Szab\'o, R., et al., 2010, MNRAS, 409, 1244
\bibitem[Szab\'o et~al.(2014)]{szabo2014} Szab\'o, R., et al., 2014, A\&A, 570, 100
\bibitem[Szab\'o et~al.(2015)]{szabo2015} Szab\'o, R., et al., 2015, EPJ WoC, 101, 01003
\end{thebibliography}
\end{document}